\documentclass[aps,preprint,floatfix]{revtex4}
\usepackage[dvips]{graphics,graphicx}
\usepackage{amsmath}
\begin{document}

\title{Bogoliubov depletion of the fragmented condensate \\in the bosonic flux ladder}
\author{Andrey~R.~Kolovsky$^{1,2}$}
\affiliation{$^1$Kirensky Institute of Physics, 660036 Krasnoyarsk, Russia}
\affiliation{$^2$Siberian Federal University, 660041 Krasnoyarsk, Russia}
\date{\today}
\begin{abstract}
We theoretically analyze  the ground state of weakly interacting bosons in the flux ladder -- the system that has been recently realized by means of ultacold atoms in the specially designed optical lattice [M. Atala, et al., Nat. Phys. 10, 588 (2014)]. It is argued that, for the system parameters corresponding to the so-called `vortex phase', the ground state is a fragmented condensate. We study the Bogoliubov depletion of this condensate and discuss the role of boundary conditions.   
\end{abstract}
\maketitle

\section{Introduction}

The possibility to mimic the bosonic flux ladder by using ultracold atoms in optical lattices, that was brought to reality in the laboratory experiment \cite{Atal14}, has renewed the theoretical interest to this unordinary quantum system \cite{Cha11,Dhar12,Hueg14,Muel14,Toku14,Pira15}. By definition, the ladder consists of two coupled one-dimensional lattices where the quantum particle hops along the ladder legs with the rate $J /\hbar$ and along the ladder rungs with the rate $J_\perp /\hbar$. The term `flux' means that the particle acquires a nonzero phase $2\pi\alpha$  when it encircles a plaquette of the ladder. For a charged particle this phase is  introduced by a magnetic field perpendicular to the ladder plane. For a neutral atom one makes the hopping matrix elements complex by using non-trivial all-optical schemes, -- the approach known nowadays as artificial or synthetic magnetic fields \cite{Aide11,Aide13,Miya13}.

One of the main features of the flux ladder is that the particle Bloch dispersion relation $E=E(\kappa)$ shows a bifurcation from that with a single minimum at zero quasimomentum $\kappa=0$ to that with two degenerate minima at $\kappa=\pm q$ as  the system parameters are varied. This bifurcation has a number of consequences for observables. In particular, the time-of-flight images of cold atoms in the flux ladder have different number of peaks depending on inequality relation between $J_\perp$ and critical $J_\perp^*=J_\perp^*(\alpha)$, as it was well observed in the cited experiment \cite{Atal14}. 

The other, perhaps even more exciting property of the system is that every Bloch state is a current-carrying state, with uniform currents flowing along the ladder legs in the opposite directions. We mention that the current is uniform only in the case of the periodic boundary condition. For other types of boundary conditions, like harmonic confinement or open boundaries, the current forms a vortex pattern with a single vortex for $J_\perp>J_\perp^*$ (single minimum in the Bloch dispersion relation) or several vortices  for  $J_\perp<J_\perp^*$ (two minima in the dispersion relation).  This change in the vortex pattern has a remote analogy with the transition from Meissner's to Abrikosov's phase in a type-II superconductors \cite{Orig01}, which is another reason explaining a large interest to the flux ladders.

Having observed a number of interesting effects already in the single-particle approximation, it is natural to ask the question about the role of atom-atom interactions which, within the Bose-Hubbard formalism, are characterized by the on-site interaction constant $U$. We recall that Bose particles in a 1D lattice with integer filling show a quantum phase transition in the parameter $U/J$ from the super-fluid phase to the Mott-insulating phase. In the past decade a lot of efforts were put towards obtaining the phase diagram of the bosonic flux ladder and the new exotic phases, like the chiral Mott insulator, were theoretically predicted in the strongly-interacting regime \cite{Dhar12}. In the present work we focus on the opposite case of weakly interacting bosons. The particular problem we address is the structure of the many-body ground state of the system  for $J_\perp<J_\perp^*$, where the single-particle Bloch dispersion relation has two degenerate minima at $\kappa=\pm q$.  

The addressed question is not trivial. In fact, let us denote by $\hat{b}^\dagger_{-q}$ and  $\hat{b}^\dagger_{+q}$ the bosonic creation operators which create the particle in the Bloch states with $\kappa=\mp q$, respectively. Then there are at least three candidates for the ground state of weakly interacting bosons:
\begin{equation}
\label{a1}
|\Psi \rangle \propto \frac{1}{\sqrt{2}}\left[e^{i\theta/2}
\frac{1}{\sqrt{N!}} \left(\hat{b}^\dagger_{+q}\right)^{N} 
+ e^{-i\theta/2}\frac{1}{\sqrt{N!}} \left(\hat{b}^\dagger_{-q}\right)^{N} 
\right] |vac\rangle \;,
\end{equation}
which is a superposition of two Bose-Einstein condensates with the relative phase $\theta$;
\begin{equation}
\label{a2}
|\Psi \rangle \propto \frac{1}{\sqrt{N!}}\left[\frac{1}{\sqrt{2}}
\left(e^{i\theta/2}\hat{b}^\dagger_{+q} + e^{-i\theta/2} \hat{b}^\dagger_{-q} \right)
\right]^{N}  |vac\rangle \;,
\end{equation}
which is a Bose-condensed state with broken translational symmetry; 
\begin{equation}
\label{a3}
|\Psi \rangle \propto \frac{1}{(N/2)!}
\left(\hat{b}^\dagger_{+q}\right)^{N/2} \left( \hat{b}^\dagger_{-q} \right)^{N/2}  |vac\rangle \;,
\end{equation}
which is a fragmented condensate where $N/2$ particles occupy the Bloch state $\psi_{-q}$ while the rest $N/2$  particles occupy the Bloch state $\psi_{+q}$. Let us also mention that Eq.~(\ref{a2}) corresponds to the mean-field ansatz, where the exact value of the phase $\theta$ is then found by minimizing the energy functional. However, the mean-field ansatz requires a justification,  which implicitly assumes the many-body analysis. This analysis  may or may not confirm the mean-field results \cite{Muel06}. For example, it is argued in Ref.~\cite{Moel10}, which discusses a relevant problem of weakly-interacting bosons in a square lattice with the uniaxial staggered flux, that the results of numerical diagonalization of the many-body Hamiltonian are consistent with the mean-field prediction (i.e., bosons do condense in a symmetry-broken state). Here we report a counterexample. It will be shown below that  the ground state of a weakly-interacting bosons in the flux ladder is the fragmented condensate (\ref{a3}). We analyze the Bogoliubov depletion of this fragmented condensate and provide an analytical estimate for the number of depleted particles. In the work we also discuss the role of boundary conditions (BC), the effect of which is particularly important for a moderate system size.

\section {Flux ladder with periodic boundary conditions}
\subsection{The model and single-particle spectrum}

We begin with the case of periodic boundary conditions where the last rung of the ladder is identified with the first rung. Labeling by $(l,m)$ the individual sites of the ladder, the system Hamiltonian reads
\begin{equation}
\label{a0}
\widehat{H}=\sum_{l=1}^L\sum_{m=1}^2
\left[-\frac{J}{2} \left(\hat{a}^\dagger_{l+1,m}\hat{a}_{l,m} + h.c.\right)
-\frac{J_\perp}{2}\left(e^{i2\pi\alpha l}\hat{a}^\dagger_{l,m+1}\hat{a}_{l,m} + h.c.\right)
+\frac{U}{2}\hat{n}_{l,m}(\hat{n}_{l,m}-1) \right] \;,
\end{equation}
where $\hat{a}^\dagger_{l,m}$ and $\hat{a}_{l,m}$ are the creation and annihilation bosonic operators and  $\hat{n}_{l,m}=\hat{a}^\dagger_{l,m}\hat{a}_{l,m}$ is the number operator. Notice that the Hamiltonian (\ref{a0}) conserves the number of particle $N$, which we assume to be much smaller than the number of sites $2L$ to avoid a transition to a Mott insulator as the interaction constant $U$ is increased. The elementary cell of the flux  ladder (\ref{a0}) is determined by the Peierls phase $\alpha$  and for a rational $\alpha=r/p$ comprises $p$ plaquettes.  Following Ref.~\cite{Atal14}, we choose $J_\perp$ to be our control parameter and we shall measure this  hopping matrix element and the interaction constant $U$ in units of the hopping matrix element $J$ (i.e., we set $J$ to unity).  

\begin{figure}[t]
\center
\includegraphics[height=8.0cm,clip]{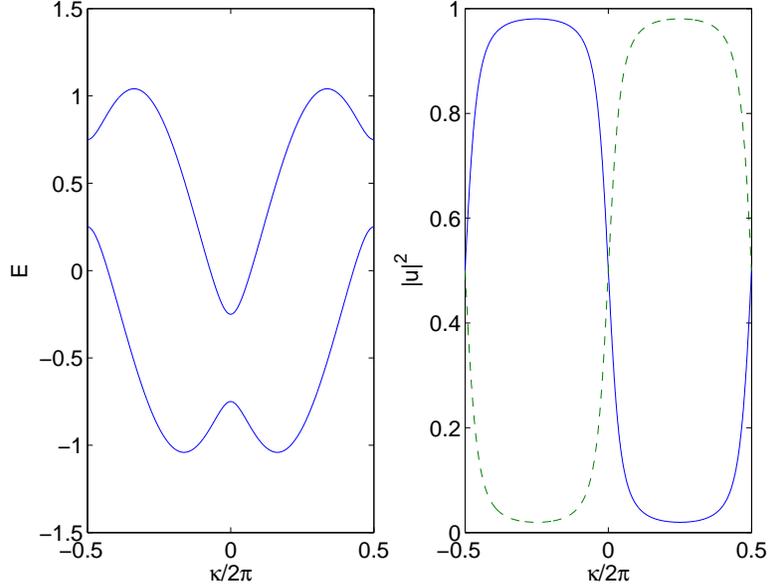}
\caption{Energy spectrum (dispersion relation) of a quantum particle in the flux ladder for $\alpha=1/3$ and $J_\perp=0.5$. The right panel shows occupation of the ladder legs for a given Bloch state, where the dashed and solid lines refer to the lower and upper leg, respectively.}
\label{fig0}
\end{figure}

The first step of analysis is to find the single-particle ($N=1$) spectrum of the system. To obtain this spectrum  it is convenient to change  the gauge as follows:
\begin{equation}
\label{a4}
\widehat{H}=\sum_{l=1}^L\sum_{m=1}^2
\left[-\frac{J}{2} \left(e^{i(-1)^m \pi\alpha}
\hat{a}^\dagger_{l+1,m}\hat{a}_{l,m} + h.c.\right)
-\frac{J_\perp}{2}\left(\hat{a}^\dagger_{l,m+1}\hat{a}_{l,m} + h.c.\right)
 \right] \;.
\end{equation}
It is seen from (\ref{a4}) that for $J_\perp=0$ the spectrum consists of two intersecting cosine dispersion relations which are shifted by $\pm \pi\alpha$ relative to $\kappa=0$. Non-zero $J_\perp$ substitutes the band crossings at $\kappa=0$ and $\kappa=\pm \pi$ by avoided crossings, thus resulting in the two-band spectrum $E=E^{(\pm)}(\kappa)$. An example is given in Fig.~\ref{fig0}(a) for $\alpha=1/3$ and  $J_\perp=0.5$, where locations of the energy minima are very close to $\kappa=\pm \pi/3$. As $J_\perp$ is increased, the minima move towards each other and eventually merge into the single minimum at $J_\perp=J_\perp^*\approx 2$. One finds a similar bifurcation of the energy spectrum for other $\alpha$, excluding the case $\alpha=1/2$, where positions of the minima are fixed at $\kappa=\pm \pi/2$.  

In addition to the dispersion relation we also need to know the explicit form of the Bloch states $\psi_\kappa$. Notice that, unlike the energy spectrum, the eigenstates do depend on the gauge.  For the Hamiltonian (\ref{a4}) the Bloch states are given by
\begin{equation}
\label{a5}
\psi^{(\pm)}_{l,m}(\kappa)=\frac{1}{\sqrt{L}}\sum_{l=1}^L e^{i\kappa l}u^{(\pm)}_m(\kappa) \;,
\end{equation}
where $u^{(\pm)}_1(\kappa)$ and $u^{(\pm)}_2(\kappa)$ are probability amplitudes to find the particle in the lower and upper leg, respectively. For the purpose of future reference the right panels in Fig.~\ref{fig0} shows squared amplitude $u_m$ as the function of the quasimomentum $\kappa$ for the ground Bloch band [minus sign in Eq.~(\ref{a5})]. It seen that for the quasimomentum $\kappa$ around the energy minimum $\kappa=+q$ the quantum particle resides mainly in the upper leg, while for $\kappa=-q$ the situation is the inverse. This strong misbalance in occupation of the ladder legs appears to be crucial for understanding the many-body spectrum of the system.

\subsection{Many-body spectrum}

\begin{figure}[b]
\center
\includegraphics[height=8.0cm,clip]{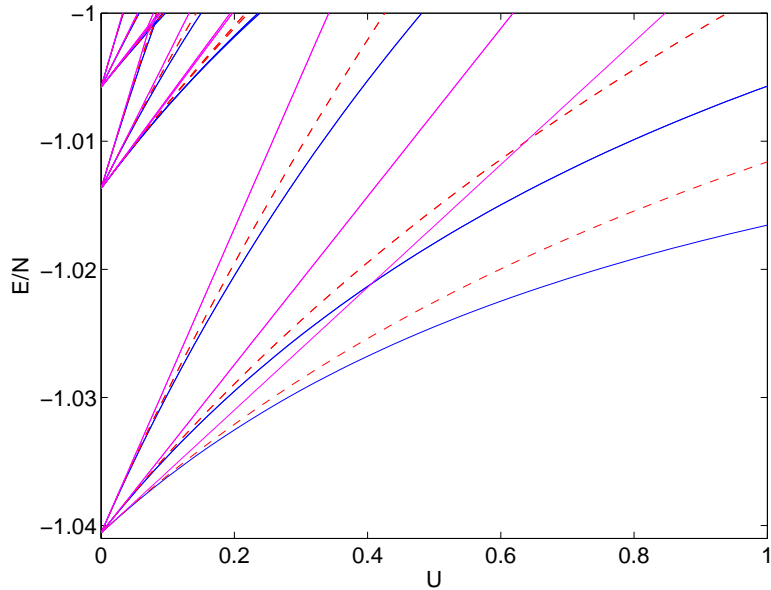}
\caption{Low-energy spectrum of $N=4$ weakly interacting bosons in the flux ladder of the length $L=12$. The other parameters are the same as in Fig.~1. Dimension of the Hilbert space ${\cal N}=17550$. Dashed lines are the spectrum in the single-band approximation, where dimension of the Hilbert space is reduced to ${\cal N}=1365$. The straight lines are the diagonal approximation.}
\label{fig1}
\end{figure}

We proceed with analysis of the many-body spectrum.  Solid lines in Fig.~\ref{fig1} show the low-energy spectrum of $N=4$ interacting bosons obtained by straightforward numerical diagonalization of the Hamiltonian (\ref{a1}) for $\alpha=1/3$, $L=12$, with the periodic BC.  Our prime interest in this figure is the lowest bunch of levels originating at $E/N=E_{min}$ where $E_{min}\approx -1.04$ is the single-particle energy  of the ground.  To analyze the depicted spectrum we shall use the approximation where one ignores the upper Bloch band. Within this approximation the low-energy spectrum of the system (\ref{a1}) is determined by the Hamiltonian 
\begin{equation}
\label{b1}
 \widetilde{H}=\sum_\kappa E(\kappa) \hat{b}_\kappa^\dagger \hat{b}_\kappa 
 + \frac{U}{2L} \sum _{\kappa_1,\kappa_2,\kappa_3,\kappa_4}
K(\kappa_1,\kappa_2,\kappa_3,\kappa_4) 
 \hat{b}_{\kappa_1}^\dagger \hat{b}_{\kappa_2}^\dagger \hat{b}_{\kappa_3} \hat{b}_{\kappa_4}
\delta(\kappa_1+\kappa_2-\kappa_3-\kappa_4) 
\end{equation}
where  $E(\kappa)=E^{(-)}(\kappa)$ is the dispersion relation for the lower band and  the kernel 
\begin{equation}
\label{b2}
K(\kappa_1,\kappa_2,\kappa_3,\kappa_4)=\sum_{m=1}^2 
u^{(-)}_m(\kappa_1) u^{(-)}_m(\kappa_2)u^{(-)}_m(\kappa_3)u^{(-)}_m(\kappa_4) \;.
\end{equation}
The spectrum calculated by using the Hamiltonian (\ref{b1}) is depicted in Fig.~\ref{fig1} by the dashed lines. A reasonable agreement is noticed. We mention, however, that the agreement becomes worse for larger $U$ or $N$ and the single-band Hamiltonian (\ref{b1}) is not suited, for example, for studying the quantum phase transition to the Mott-insulator state.
   
Next we employ the perturbation theory. Let us denote by $|\Psi_m \rangle$ the degenerate ground states of the system (\ref{b1}) for $U=0$,
\begin{equation}
\label{a6}
|\Psi_m \rangle =|\ldots,N/2-m,\ldots,N/2+m,\ldots\rangle \;,\quad |m| \le N/2 \;.
\end{equation}
The states (\ref{a6}) obviously correspond to $2N+1$ possible distribution of $N$ bosons between two energy minima. Infinitesimally small $U$ removes the degeneracy where, according to the first-order perturbation theory, $E_m(U)=\langle \Psi_m|\widetilde{H}|\Psi_m\rangle$. After some algebra this gives
\begin{equation}
\label{b3}
E_m(U)=E_{min}N+\frac{U}{2L}[K_1( N^2/2+2m^2-N)+K_2(N^2-4m^2)] \;,
\end{equation}
where $K_1=K(q,q,q,q)$ and $K_2=K(q,-q,q,-q)$.  The first term in the square brackets in Eq.~(\ref{b3}) is the self-energy of the state (\ref{a6}) due to virtual annihilation and creation of the particles in the same quasimomentum state [i.e., $\kappa_1=\kappa_2=\kappa_3=\kappa_4$ in Eq.~(\ref{b1})]. The second term in the square brackets is a correction  to the self-energy due to virtual annihilation and creation of the particles in the different quasimomentum states $\kappa=-q$ and $\kappa=+q$. Since the single-particle wave function $\psi_{l,m}(\kappa=\pm q)$ is essentially localized in one of two legs [see Fig.~\ref{fig0}(b)],  we have $K_2\ll K_1$. For example, for the considered ratio $J_\perp/J=0.5$ the parameter $K_1= 0.95$ and $K_2=0.05$. 

The presented elementary analysis brings us to the first important conclusion. It follows from Eq.~(\ref{b3}) that the lowest level in Fig.~\ref{fig1} corresponds to $m=0$ and, hence, the state $|\Psi_0\rangle$,  which is a fragmented condensate with $N/2$ particles occupying the quasimomentum state $\kappa=-q$ and the rest $N/2$ particles occupying the state $\kappa=+q$, provides a zero-order approximation to the exact ground state. To find the latter analytically one should take into account the Bogoliubov depletion of the fragmented condensate $|\Psi_0\rangle$, i.e., non-zero occupations of the other quasimomentum states besides $\kappa=\pm q$.  The Bogoliubov depletion is known to lower the system energy as compared to the diagonal approximation, and even a small depletion can cause considerable deviation from the linear law (\ref{b3}). For example, for the parameter of Fig.~\ref{fig1} and $U=1$ the relative depletion of the fragmented condensate is $N_d/N=0.0816$. In the next subsection we discuss the Bogoliubov depletion of the fragmented condensate in more details.

\subsection{Bogoliubov depletion of the fragmented condensate}

To find occupation probabilities of the other quasimomentum states we shall use the following ansatz:
\begin{equation}
\label{b5}
|\Psi\rangle=\sum_{n,k} c_{n,k} |\ldots,n,N/2-2n,n,\dots,k,N/2-2k,k,\ldots\rangle \;.
\end{equation}
Physically, the ansatz (\ref{b5}) corresponds to the process where one `takes' two particles from either of condensates and put them symmetrically  (to satisfy the conservation law for the total quasimomentum) in the nearest quasimomentum states.  Substituting (\ref{b5}) into the stationary Schr\"odinger equation with the Hamiltonian (\ref{b1}) we obtain the following equation on the coefficients $c_{n,k}$, 
\begin{equation}
\label{b7}
E_0 c_{n,k}+\delta (n+k) c_{n,k} + \left( V^{(-q)}_{n,k} c_{n+1,k} + h.c. \right)
+\left( V^{(+q)}_{n,k} c_{n,k+1} + h.c. \right) = E c_{n,k} \;,
\end{equation}
where $E_0=E_{min}N+(U/2L)K_1(N^2/2-N)$  is the energy of not-depleted fragmented condensate and $\delta$,
\begin{equation}
\label{bc3}
2\delta=E(\pm q-\Delta\kappa)+E(\pm q+\Delta\kappa)-2E_{min}  \;,
\end{equation}
is an increase in the kinetic energy due to population of the nearest quasimomentum states $\kappa=-q \pm \Delta\kappa$ and  $\kappa=+q \pm \Delta\kappa$.

\begin{figure}
\center
\includegraphics[height=8.0cm,clip]{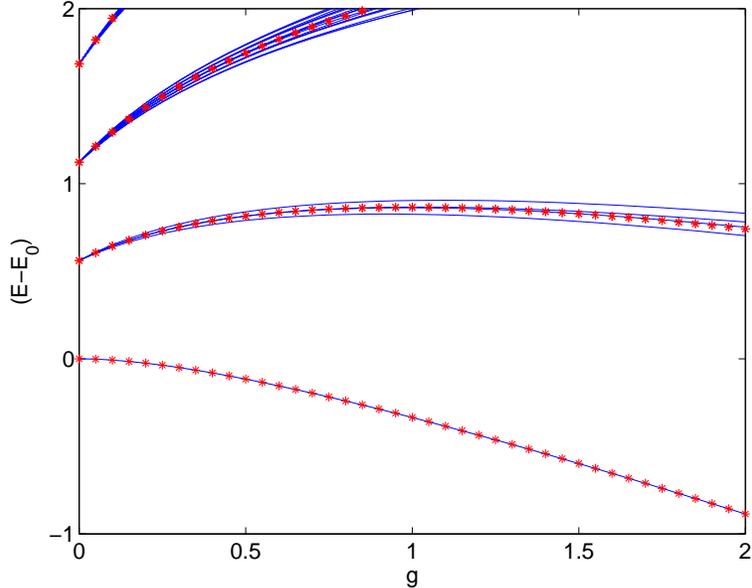}
\caption{Low-energy spectrum of $N=512$ bosons in the flux ladder ($\alpha=1/2$) of the length $L=8$ as the function of the macroscopic interaction constant $g$. The energy is measured with respect to the mean-field energy $E_0$. Asterisks are the spectrum calculated by using the ansatz (\ref{b5}).}
\label{fig2}
\end{figure}

Next  we shall assume the semiclassical limit  $N\rightarrow\infty$ and $U=gL/N\rightarrow0$, where 
\begin{equation}
\label{b4}
g=NU/L 
\end{equation}
is the macroscopic interaction constant, In this limit the off-diagonal matrix elements  $V^{(\pm q)}_{n,k}$ take a particular simple form, namely, $V^{(\pm q)}_{n,k}=K_1g/2$.  Furthermore, in the semiclassical limit Eq.~(\ref{b7}) separates into two independent eigenvalue problems, so that $c_{n,k}=c_n c_k$.  The corresponding equations for the coefficients $c_n$ and $c_k$ are identical and read
\begin{equation}
\label{b8}
2(\delta+\tilde{g})n c_n+\tilde{g}(n c_{n+1} +h.c.)=(E-E_0)c_n  \;,\quad E_0\approx E_{min}N+\tilde{g}N/4 \;, \quad \tilde{g}=K_1g/2 \;.
\end{equation}
Eq.~(\ref{b8}) can solved analytically, resulting in the equidistant spectrum \cite{70},
\begin{equation}
\label{b6}
E_j=E_0-(\delta+\tilde{g})+\Omega(j+1/2) \;,\quad \Omega=2\sqrt{2\tilde{g}\delta+\delta^2} \;,
\end{equation}
which is nothing else as the Bogoliubov excitations of a Bose-Einstein condensate. For the purpose of future comparison asterisks in Fig.~\ref{fig2} show the energy spectrum of the system (\ref{b1})  calculated on the basis of Eq.~(\ref{b6}) for the system parameters $N=512$, $L=8$, and $\alpha=1/2$ .

The approximation (\ref{b5}) for the low-energy eigenstates can be improved further by taking into account the `correlated' depletion process, which corresponds to the following ansatz,
\begin{equation}
\label{bc2}
|\Psi\rangle=\sum_{n,k} c_{n,k} |\ldots,n,N/2-n-k,k,\dots,k,N/2-n-k,n,\ldots\rangle \;.
\end{equation}
In the other words,  we take one particle from each condensates and distribute them symmetrically between the nearest quasimomentum states. We found that the depletion process (\ref{bc2}) corrects Eq.~(\ref{b6}) only slightly. Yet it removes degeneracy of the excited states $E_j$, where the level splitting  is proportional to the parameter $K_2$ (see solid lines in Fig.~\ref{fig2}).  Notice that, unlike in Fig.~\ref{fig1}, in Fig.~\ref{fig2} we depict only the energy levels which correspond to zero total quasimomentum. 

The presented results prove that, providing the condition $K_1\gg K_2$, the Bogoliubov depletion of the fragmented condensate is similar to that of not-fragmented condensate. This allows us to obtain an analytical estimate for the number of depleted particles. Namely,  knowing the eigenstates of Eq.~(\ref{b8}) we calculate the depletion of each condensate as $N_d=\sum 2n|c_n|^2$. For the ground state of the original system this gives
\begin{equation}
\label{b9}
N_d\approx 2\sqrt{\tilde{g}/\delta} \;, \quad \tilde{g}=K_1g/2 \;.
\end{equation}
It follows from Eq.~(\ref{b9}) that $N_d$ diverges in the thermodynamic limit $N,L\rightarrow\infty$,  where the parameter $\delta$ tends to zero as $\delta\sim 1/L^2$. This is in agreement with the well-known result that there could be no genuine condensate in an infinite 1D system. Fortunately, in laboratory experiments with cold atoms one always deals with finite systems and, thus, $N_d$ remains finite.

\subsection{One-particle density matrix}

\begin{figure}
\center
\includegraphics[height=8.0cm,clip]{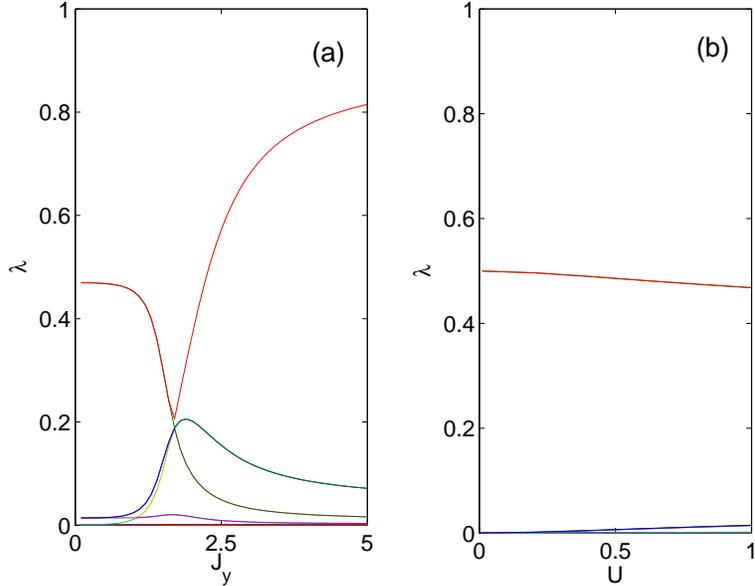}
\caption{Eigenvalues of the one-particle density matrix (\ref{a7}) as the function of the hopping matrix element $J_\perp$, left panel, and as the function of the interaction constant $U$, right panel. The other parameters are  $N=4$, $L=12$, $\alpha=1/3$, and $U=1$ (left panel) and $J_\perp=0.5$ (right panel). }
\label{fig3}
\end{figure}

The above discussed depletion of the fragmented condensate can be easily detected  by analyzing the one-particle density matrix,
\begin{equation}
\label{a7}
\rho_{l,m}^{l',m'}=\frac{1}{N} \langle\Psi | \hat{a}^\dagger_{l',m'} \hat{a}_{l,m} | \Psi \rangle  \;,
\end{equation}
where $|\Psi\rangle$ is the ground state of the system (\ref{a0}).  The right panel in Fig.~\ref{fig3} shows eigenvalues $\lambda_i$ of the matrix (\ref{a7}),
\begin{equation}
\label{a8}
\rho_{l,m}^{l',m'}=\sum_{i=1}^{2L} \lambda_i  \phi_{l',m'}^{*(i)}\phi_{l,m}^{(i)}  \;,
\end{equation}
as the function of the interaction constant $U$. It is seen that the density matrix has two degenerate eigenvalues close to 1/2, that correspond to macroscopic occupations of the Bloch states with $\kappa=\pm q$. A degrease of these two largest eigenvalues as the interaction strength is increased obviously indicates the depletion of the fragmented condensate. 

It is interesting to note  that the system of a rather small size can qualitatively reproduces the quantum phase transition  associated with the bifurcation of the single-particle energy spectrum at $J_\perp^*$.  The left panel in Fig.~\ref{fig3} shows eigenvalues of the density matrix as the function of $J_\perp$. It is seen in Fig.~\ref{fig3}(a) that the fragmented condensate rapidly depletes as $J_\perp$ approaches the critical value $J_\perp^*$ and then an ordinary condensate, where bosons condense in the Bloch state with zero quasimomentum, is formed.

To conclude this subsection we briefly comment on the time-of-flight (TOF) images of the fragmented condensate. In principle, the knowledge of the one-particle density matrix (\ref{a8}) suffices to predict TOF images  of the expanding fragmented condensate.  However, these predictions are valid only the statistical sense. In the other words, the TOF images should be overaged over many runs of the identical experiment. At the single run the TOF image is expected to be similar to that for not-fragmented condensate in the form (\ref{a2}), where one randomly peaks up the phase $\theta$. This conclusion formally follows from the fact that the translationally invariant state (\ref{a3}) can be obtained from the symmetry-broken state (\ref{a2}) by averaging the latter over the phase $\theta$ \cite{remark2}.

\section{Dirichlet boundary condition} 
\subsection{Single-particle ground states}

As mentioned in the Introduction, the Dirichlet BC drastically changes the structure of the single-particle wave functions and, instead of uniform currents in the ladder legs, we get a vortex pattern for the persistent current. If $J_\perp>J_\perp^*$ there is only one vortex extending over the system size, while for $J_\perp<J_\perp^*$ there are several vortices with the characteristic size $\pi/|q|$.  As before, we focus on the latter case where the Bloch dispersion relation for the periodic BC condition would have two degenerate minima.  

\begin{figure}
\center
\includegraphics[height=9cm,clip]{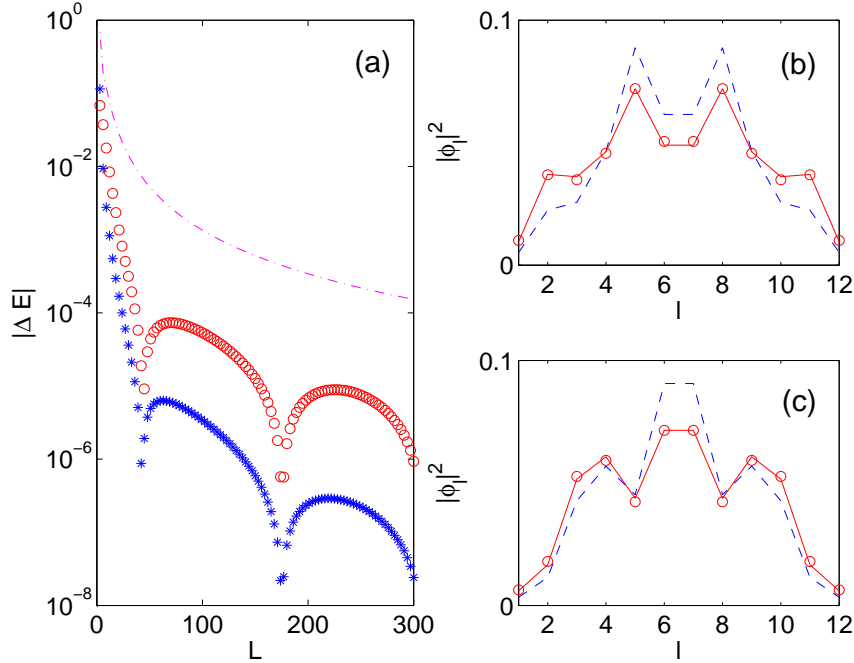}
\caption{Panel (a): The energy difference $|\Delta E|$ between two lowest eigenstates for $g=0$, asterisk, and $g=1$, open circles. The dashed line is the energy difference between the lowest duplet and the next eigenstate. The considered system size $3\le L\le 300$ with the step $\Delta L=3$. 
Panels (b) and (c): Symmetric and antisymmetric ground states for $N=4$, $L=12$ and $U=0$, dashed lines, $U=1$, solid lines. Shown are the squared amplitudes $|\tilde{\psi}_{l,m}|^2$ and $|\phi_{l,m}|^2$, which are the same for $m=1$ and $m=2$.} 
\label{fig45}
\end{figure}

For the Dirichlet  BC the new ground states of the system can be approximately expressed through the Bloch states (\ref{a5}) as  
\begin{equation}
\label{c1}
\tilde{\psi}_{l,m}^{(\pm}= Y(l)\frac{1}{\sqrt{2}}[\psi_{l,m}(+q) \pm \psi_{l,m}(-q)]     \;, 
\end{equation}
where $Y(l)\approx L^{-1/2} \sin(\pi l/L)$ is a smooth envelope function. Examples of the symmetric (plus sign) and antisymmetric (minus sign) ground states are given by the dashed lines in the  panel (b) and  (c) in Fig.~\ref{fig45} for $L=12$ and in Fig.~\ref{fig44} for $L=120$. Notice that, unlike the Bloch states for periodic BC, the states (\ref{c1}) have equal population of the ladder legs, i.e., $|\tilde{\psi}_{l,1}|^2=|\tilde{\psi}_{l,2}|^2$. Furthermore, the states (\ref{c1}) obey the following symmetry,
\begin{equation}
\label{c7}
\tilde{\psi}_{l,1}^{(\pm)} = \pm \tilde{\psi}_{L-l+1,2}^{(\pm)}  \;,
\end{equation}
which we shall utilize later on. (We mention, in passing, that one finds a similar symmetry also in the case of harmonic confinement, which is a more relevant BC from the viewpoint of a laboratory experiment.)

The other important result of the single-particle analysis is that the new  ground states (\ref{c1}),  which are locally given by symmetric and antisymmetric superpositions of the Bloch waves with the quasimomentum  $\kappa=\pm q$,  have slightly different energies. The energy difference $\Delta E =|E^{(-)}-E^{(+)}|$ between the symmetric and antisymmetric states can be very small, yet it remains finite. This is illustrated in Fig.~\ref{fig45}(a) where we depict by asterisks  the energy difference $\Delta E$ as the function of the system size $L$.

\subsection{Many-body ground state}

\begin{figure}
\center
\includegraphics[height=8.0cm,clip]{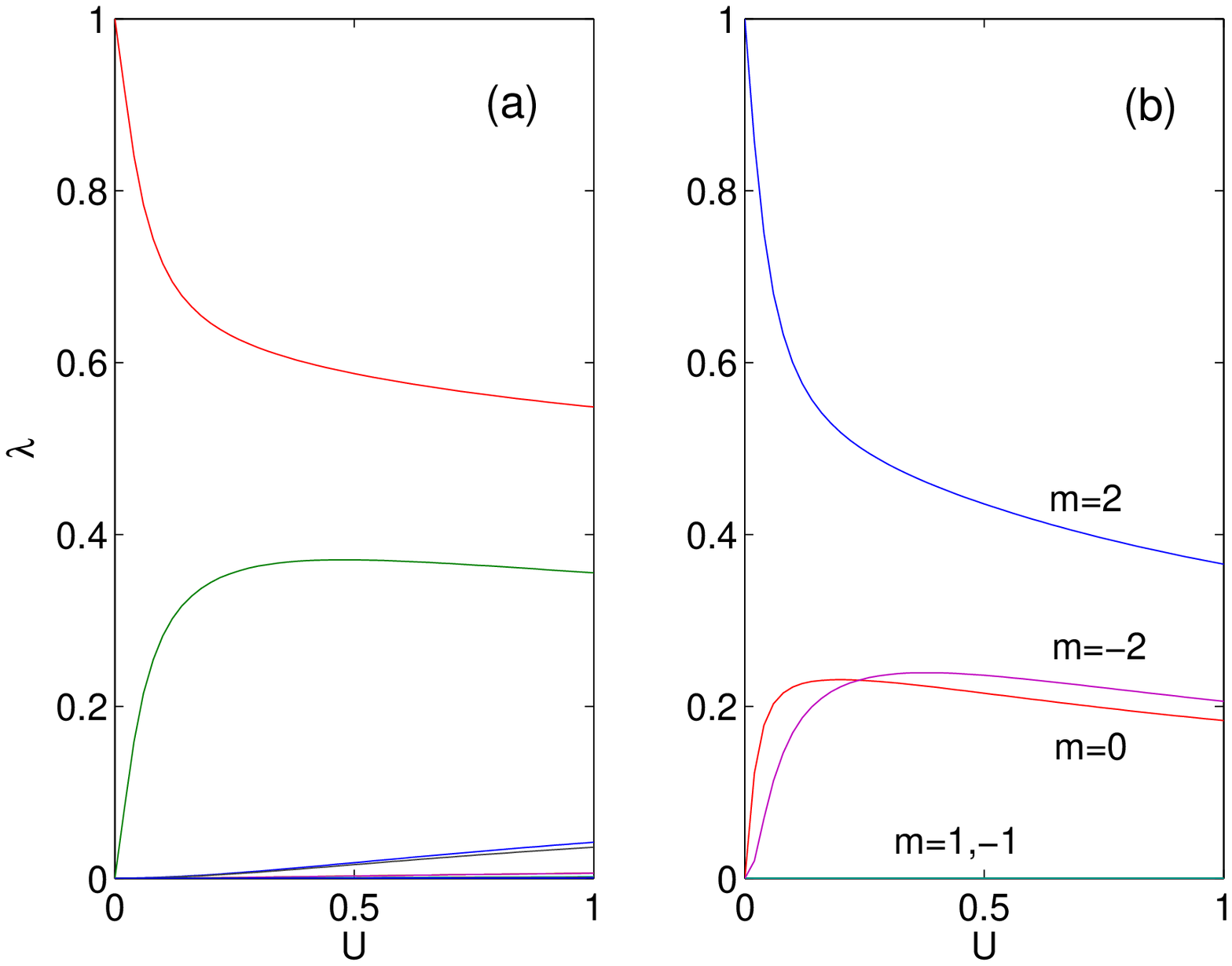}
\caption{Eigenvalues of the one-particle density matrix (\ref{a7}), left panel, and the overlap coefficients (\ref{c2}), right panel, as the function of the interaction constant $U$. The system parameters are the same as in Fig.~\ref{fig3}(b) except the different (Diriclhet) boundary condition.}
\label{fig4}
\end{figure}

It follows from the previous subsection that for vanishing inter-particle interactions the many-body ground state is an ordinary condensate, where bosons occupy the symmetric single-particle state. However, since the energy difference $\Delta E$ between the symmetric and antisymmetric states is rather small, one may expect that interactions force bosons to occupy both single-particle states. This conclusion is supported by the numerical analysis of the problem -- see Fig.~\ref{fig4}(a), which show eigenvalues of the density matrix (\ref{a7}) as the functions of the interaction constant $U$. As $U$ is increased a considerable fraction of particles is seen to populate the antisymmetric state  $\tilde{\psi}_{l,m}^{(-)}$, which has a slightly larger energy $E^{(-)}=E^{(+)}+ 0.0011$. Additionally, the solid lines in Fig.~\ref{fig45}(b,c) show the eigenvectors $\phi_{l,m}$   of the density matrix for two largest eigenvalues, which should be compared with the single-particle states $\tilde{\psi}_{l,m}^{(\pm)}$ depicted by the dashed lines.

To obtain further information about the many-body ground state we project the ground state wave function onto the states $|\Psi_m\rangle$,
\begin{equation}
\label{c8}
|\Psi_m \rangle = \frac{1}{\sqrt{(N/2+m)!(N/2-m)!}}
\left(\tilde{b}^\dagger_{+}\right)^{N/2+m} \left( \tilde{b}^\dagger_{-} \right)^{N/2-m}  |vac\rangle \;,
\end{equation}
where the operators $\tilde{b}^\dagger_{\pm}$ creates the particle in the symmetric and antisymmetric single-particle states, respectively.  The result is shown in the right panel in Fig.~\ref{fig4}, where we plot squared modulus of the overlap coefficients 
\begin{equation}
\label{c2}
C_m=\langle \Psi| \Psi_m\rangle \;.
\end{equation}
It is seen in Fig.~\ref{fig4}(b) that the exact ground state $|\Psi\rangle$ is a sum of three condensates, one fragmented and two not-fragmented. Notice that strictly zero overlap coefficients $C_{\pm 1}$ prove that this state cannot be viewed as a single condensate, where bosons are condensed in a some imbalanced superposition of the symmetric and antisymmetric states.

The above presented results suggest the following physical picture. For a large system size, where the energy difference $\Delta E$ between the symmetric and antisymmetric single-particle states is negligible, the ground state is the balanced fragmented condensate, similar to the case of periodic BC. As the system size is decreased, the energy difference $\Delta E$ increases and we have admixture of not-fragmented condensates.  We mention that one meets a similar effect in the problem of a single-mode spinor condensate subject to a magnetic field \cite{Muel06}.  In both problems we have the energy difference between the single-particle states as an additional parameter. It is an open problem to find the overlap coefficients (\ref{c2}) as the function of the energy splitting $\Delta E$ or, what is essentially the same, as the function of the system size $L$.

\subsection{Mean-field macroscopic wave-functions}

Finally, we analyze the macroscopic wave functions $\phi_{l,m}^{(\pm)}$ given by the first two eigenstates of the one-particle density matrix.  We shall assume the system size to be large, so that the ground state of the system is a balanced fragmented condensate, and inter-particle interactions is weak, so that the Bogoliubov depletion of this condensate is small.  In this case we can find the functions $\phi_{l,m}^{(\pm)}$ by using  the mean-field approach:
\begin{equation}
\label{c4}
-\frac{J}{2}\left(\phi_{l+1,m} e^{-i\pi\alpha} + \phi_{l-1,m} e^{i\pi\alpha} \right)
-\frac{J_\perp}{2}\left(\phi_{l,m+1} \delta_{m,1}+ \phi_{l,m-1}\delta_{m,2} \right)
+g|\phi_{l,m}|^2 \phi_{l,m} = E \phi_{l,m} \;,
\end{equation}
where the macroscopic interaction constant $g$ is defined in Eq.~(\ref{b4}). The stationary nonlinear Schr\"odinger equation (\ref{c4}) can be simplified by noting that the ground single-particle states (\ref{c1}) posses the symmetry (\ref{c7}) and so do the eigenstates  $\phi_{l,m}^{(\pm)}$ of the one-particle density matrix. Using this symmetry Eq.~(\ref{c4}) reduces to two one-dimensional problems,
\begin{equation}
\label{c5}
-\frac{J}{2}\left(\phi_{l+1} e^{-i\pi\alpha} + \phi_{l-1}e^{i\pi\alpha} \right)
\pm \frac{J_\perp}{2}\phi_{L-l+1}  + 2g|\phi_{l}|^2 \phi_{l} = E \phi_{l} \;,
\end{equation}
%
\begin{figure}[t]
\center
\includegraphics[height=8.5cm,clip]{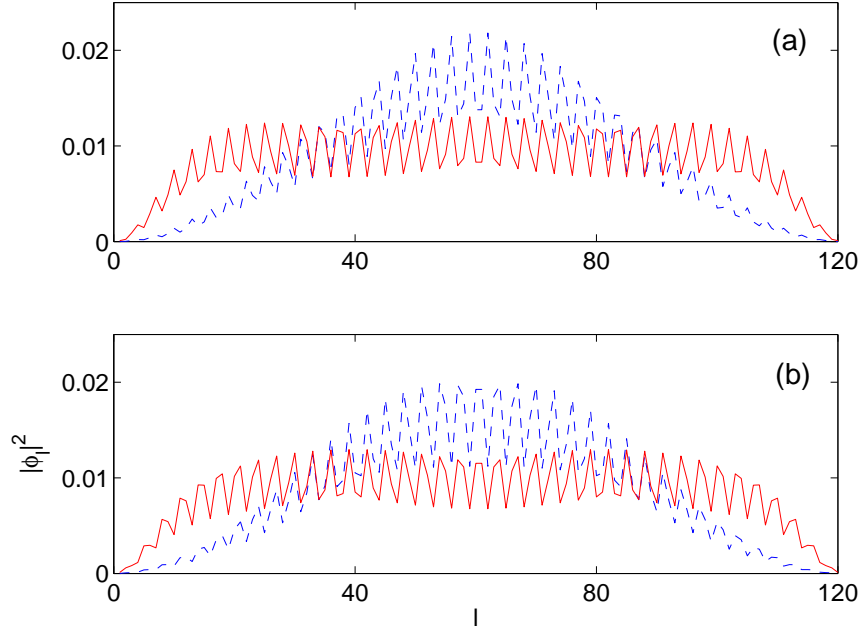}
\caption{Occupations $|\phi_{l,m}|^2$ of the ladder sites for $L=120$ and $g=0$, dashed lines, and $g=1$, solid lines, for the symmetric, upper panel, and antisymmetric, lower panel, macroscopic wave functions.}
\label{fig44}
\end{figure}
where the plus and minus sign refer to the symmetric and antisymmetric macroscopic wave functions, respectively. For $L=120$ and $g=1$ the minimal energy solutions of Eq.~(\ref{c5}) are depicted in Fig.~\ref{fig44} by the solid lines.  It is seen that repulsive interactions make the envelope function $Y(l)$ flatter, which is consistent with the role of interactions in the case of not uniform wave functions. Yet, there are unremovable  modulations of the particle density with the period $\pi/|q|\approx 3$, which are correlated with the vortex pattern for the persistent current. For the sake of completeness we also depict in Fig.~\ref{fig45}(a) the difference $\Delta E$ between the mean-field energies of the symmetric and antisymmetric states.

It is interesting to note that, if we sum  $|\phi_{l,m}^{(+)}|^2$ and $|\phi_{l,m}^{(-)}|^2$ with equal weights, the density becomes locally uniform. Keeping in mind that the uniform density minimizes the the mean-field interaction energy, this result supports the above conclusion that in the limit of large $L$ the ground state of the system is the balanced fragmented condensate.

\section{Conclusion}

In the work we discussed the ground state of weakly interacting bosons in the flux ladder, focussing on the case where the single-particle Bloch spectrum has two degenerate minima at $\kappa=\pm q$. It was shown that the many-body ground state is the fragmented condensate, where $N/2$ particles occupy the Bloch state with the quasimomentum $\kappa=-q$ and the rest $N/2$ particle the Bloch state with the quasimomentum $\kappa=+q$. We analyzed the Bogoliubov depletion of this fragmented condensate and obtained an analytical estimate for the number of depleted particles as the function of the interaction constant and the system size. 

In the second part of the paper we addressed the case of Dirichlet BC instead of the periodic BC. The Dirichlet BC  explicitly break  the translational symmetry of the system, so that the new nearly degenerate single-particle states shows the vortex pattern for the persistent current. It was argued that, in the limit of large system size, the ground state of the system is again the fragmented condensate where the particles occupy the `symmetric'  and `antisymmetric' macroscopic states. We calculated these macroscopic states by using a specific mean-field approach. This approach explicitly  take into account the structure of the exact many-body ground state and cardinally  differs from the standard mean-field approach, where the ground state is assumed to be an ordinary (not-fragmented) condensate.   

{\em Acknowledgments}. The author acknowledges discussions with A. Eckardt, hospitality of MPIPKS in Dresden, and financial support from RFBR and KFN through the grant No. 16-42-240746.



\begin{thebibliography}{99}
\bibitem{remark2}
A rigorous proof of the above statement  would require a deeper insight into the measurement process,  in spirit of the work \cite{Java96} aimed to explain the results of the fundamental experiment  \cite{Andr97} with fragmented Bose-Einstein condensate in a double-well potential.
\end{thebibliography}
\end{document}